\documentclass[12pt]{article}
\usepackage{epsfig,amssymb}
\usepackage{graphicx,color}

\def\appendix{{\newpage\section*{Appendix}}\let\appendix\section%
        {\setcounter{section}{0}
        \gdef\thesection{\Alph{section}}}\section}

\newcommand{\be}{\begin{equation}}
\newcommand{\ee}{\end{equation}}
\newcommand{\bear}{\begin{eqnarray}}
\newcommand{\eear}{\end{eqnarray}}
\newcommand{\ba}{\begin{array}}
\newcommand{\ea}{\end{array}}
\newcommand{\nn}{\nonumber}

\begin{document}

\begin{flushright}
{\tt hep-th/yymmnnn}
\end{flushright}
\vspace{5mm}

\begin{center}
{{\Large \bf 3-dimensional holographic trace anomaly\\
from AdS/CFT correspondence}\\[14mm]
{Chong Oh Lee}\\[2.5mm]
{\it Department of Physics,\\
Chonbuk National University, Jeonju 561-756, Republic of Korea}\\
{\tt cohlee@chonbuk.ac.kr}\\[4mm]
{Chaiho Rim}\\
{\it Department of Physics and Research Institute of Physics and Chemistry,
\\ Chonbuk National University,
Jeonju 561-756, Republic of Korea}\\
{\tt rim@chonbuk.ac.kr} }
\end{center}

\vspace{10mm}

\begin{abstract}
We explicitly obtain energy-momentum tensor at the asymptotic
3-dimensional region of Schwarzschild AdS$_4$ and Taub-NUT-(A)dS$_4$
using the so-called `counter-term subtraction method'
in Fefferman-Graham coordinate.
The energy momentum tensor is presented in a closed form
for the AdS$_4$ and for the special case of Taub-NUT-dS
and in an asymptotic series for other cases.
The result suggests that in light of AdS/CFT correspondence,
the 3-dimensional trace anomaly can be  expressed
in terms of the 3-dimensional volume and Ricci scalar.
\end{abstract}

\newpage
\setcounter{equation}{0}
\section{Introduction}
The AdS/CFT correspondence conjecture
\cite{Maldacena:1997re,Gubser:1998bc,Witten:1998qj,Aharony:1999ti,D'Hoker:2002aw}
asserts there is an
equivalence between a gravitational theory in the bulk and a conformal field theory in the
boundary.
The correspondence is shown to hold for AdS \cite{Fefferman:1984aa}, S-AdS
\cite{Balasubramanian:1999re,Emparan:1999pm,Kraus:1999di}, Taub-NUT-AdS
\cite{Chamblin:1998pz,Hawking:1998ct,Mann:1999bt} and for
the generalized version of Taub-NUT-AdS$_4$ \cite{Clarkson:2002uj,Astefanesei:2004kn}
whose metric is suggested by \cite{Awad:2000gg}.

The AdS action is given as
\bear\label{totadsact}
A=A_B+A_{\partial B}+A_{ct},
\eear
where $A_B$ is  the bulk action $A$ in $d+1$-dimensional Manifold $\cal M$
and $A_{\partial B}$ is Gibbons-Hawking action,
\bear\label{adsact}
A_B&=&\frac{1}{2k^2}\int_{\cal M}d^{d+1}x\sqrt{-G}({\cal R}-2\Lambda),\nonumber\\
A_{\partial B}&=&-\frac{1}{k^2}\int_{\partial {\cal M}}d^d x
\sqrt{-\gamma}\Theta\,.
\eear
Here $k^2=8\pi G_{d+1}$ with
$d+1$-dimensional gravitational constant $G_{d+1}$,
$\Lambda=-{d(d-1)}/{2l^2}$ is a negative cosmological constant
and $\Theta$ is the trace of extrinsic curvature.
The boundary action $A_{\partial B}$ is added to make
equations of motion well behaved at the boundary
and boundary energy-momentum (BEM) tensor is expressed by \cite{Brown:1992br}
\bear\label{bdryenergymoment}
\frac{2}{\sqrt{-\gamma}}\frac{\delta A_{\partial B}}{\delta \gamma^{ab}}=\Theta_{ab}
-\gamma_{ab}\Theta\,.
\eear

The counter-term action $A_{ct}$ is added to remove
the infrared divergence which appears
when the boundary goes to infinity.
The first few terms were explicitly evaluated in
\cite{Balasubramanian:1999re,Emparan:1999pm}
and the universality of the divergent structure
determines the counter-term action $A_{ct}$ in arbitrary dimensions
\cite{Kraus:1999di}
\bear\label{bdryenergymomentct}
A_{ct}&=&-\frac{1}{k^2}\int_{\partial {\cal M}} d^d x \sqrt{-\gamma}
\left\{-\frac{d-1}{l}-\frac{l R}{2(d-2)}{\cal F}(d-3) \right.\nonumber\\
&&-\frac{l^3}{2(d-2)^2(d-4)}\left(R_{ab}R^{ab}-\frac{d}{4(d-1)}R^2\right)
{\cal F}(d-5)\nonumber\\
&&+\frac{l^5}{(d-2)^3(d-4)(d-6)}\left(\frac{3d+2}{4(d-1)}RR_{ab}R^{ab}
-\frac{d(d+2)}{16(d-1)^2}R^3\right.\nonumber\\
&&\left.\left.
+\frac{d-2}{2(d-1)}R^{ab}\nabla_a \nabla_b R -R^{ab}\Box R_{ab}+\frac{1}{2(d-1)}R\Box R
\right){\cal F}(d-7)+\cdots\right\},
\eear
where ${\cal F}(d)$ is a step function, 1 when $d\geq 0$, 0 otherwise.

The counter-term action $A_{ct}$ is best understood in
Fefferman-Graham (FG) coordinate. The AdS metric near the boundary
is given as \cite{Fefferman:1984aa}
\bear\label{FGseries}
ds^2=G_{\mu\nu}dx^{\mu}dx^{\nu}=\frac{l^2}{z^2}dz^2+\frac{l^2}{z^2}
g_{ab}(x, z)dx^a dx^b,
\eear
where the $r=\infty$ is put to $z=0$.
The Greek indices $\mu$ and $\nu$ refer to volume coordinates, $1,
\cdots, d+1$ and the roman $a$ and $b$ to boundary coordinates,
$1,\cdots,d$ and $\gamma_{ab}$ in (\ref{bdryenergymoment}) is given
as $\frac{l^2}{z^2}g_{ab}$. The Einstein equation for the bulk
metric has two different types of asymptotic forms for most of
models (see counterexample in \cite{Berg:2001ty}); when the boundary
dimensions $d$ is odd
\bear\label{FGbodaryserieseven}
g_{ab}(x,z)=\sum_{p=0}^{(d-1)/2}g_{ab}^{(2p)}(x)z^{2p}+g_{ab}^{(d)}(x)z^d+{\cal
O}(z^{d+2}),
\eear
and when $d$ is even
\bear\label{FGbodaryseries}
g_{ab}(x,z)=\sum_{p=0}^{d/2}g_{ab}^{(2p)}(x)z^{2p}+h_{ab}^{(d)}(x)z^d
\ln z^2+ {\cal O}(z^{d+2}),
\eear
where $g_{ab}^{(j)}(x)$ and
$h_{ab}^{(j)}(x)$ contain $j$ number of derivatives with respect to
$x^a$. This shows that the odd powers of $z$ do not appear in
$g_{ab}(x,z)$ up to the order of $z^d$. Explicit calculation has been done
for AdS$_5$ in \cite{Nojiri:2002td,Lee:2007pb}. This also fits for
Schwarzschild AdS (S-AdS$_4$) and Taub-NUT-(A)dS$_4$ as shown in the
next sections.

Dimensional analysis of the extrinsic curvature
allows the terms of order $p<d/2$ only
in (\ref{FGbodaryserieseven}) and (\ref{FGbodaryseries})
to the divergent part of the action \cite{Balasubramanian:1999re}.
Therefore, the divergent terms of the BEM tensor is given as
\bear\label{bdryemexpansion}
\tilde{T}_{ab} = \sum_{p=0}^{[(d-1)/2]}\tilde{T}_{ab}^{(p)},
\eear
where $[x]$ is the Gauss number (greatest integer less than or equal to $x$).
$\tilde{T}_{ab}^{(p)}$ has the power of $l^{2p-1}$ and is
determined through the Gauss-Codazzi equations \cite{Kraus:1999di,Wald:1984aa},
\bear\label{constraneq}
\frac{1}{d-1}\tilde{T}^2-\tilde{T}_{ab}\tilde{T}^{ab}=\frac{d(d-1)}{l^2}+R.
\eear
Noting that $\tilde{T}_{ab}$ is derived from the counter-term action
\bear\label{counterenergymo}
\tilde{T}_{ab} =\frac{2}{\sqrt{-\gamma}} \frac{\delta A_{ct}} {\delta \gamma^{ab}}
=\frac{2}{\sqrt{-\gamma}} \frac{\delta} {\delta \gamma^{ab}}\int d^d x
\sqrt{-\gamma} {L}_{ct},
\eear
and the counter-term action should be invariant under the local Weyl variations
$\delta_{W}\gamma_{ab}=\sigma\gamma_{ab}$
according to the AdS/CFT conjecture, one has
\cite{Kraus:1999di}
\bear\label{conunterlagran}
(d-2p){L}_{ct}^{(p)}=\tilde{T}^{(p)}_{ab}\gamma^{ab}.
\eear
Thus the full BEM tensor $T_{ab}$ is given by
\bear\label{fullemt}
T_{ab}&=&\frac{2}{\sqrt{-\gamma}}\frac{\delta (A_{\partial B}+A_{ct})}{\delta \gamma^{ab}}
=\frac{1}{8\pi G_{d+1}}\biggr[\Theta_{ab}-\Theta\gamma_{ab}\biggr]
+\sum_{p=0}^{[(d-1)/2]}\tilde{T}_{ab}^{(p)},
\eear
where a few terms of $\tilde{T}_{ab}^{(p)}$'s are explicitly given by
\bear\label{emts}
&&\tilde{T}_{ab}^{(0)}=-\frac{d-1}{l}\gamma_{ab},\nonumber\\
&&\tilde{T}_{ab}^{(1)}=\frac{l}{d-2}\left(R_{ab}-\frac{1}{2}\gamma_{ab}R\right){\cal F}(d-3),
\nonumber\\
&&\tilde{T}_{ab}^{(2)}=\frac{l^3}{(d-2)^2(d-4)}\left\{-\frac{1}{2}\gamma_{ab}
\left(R_{cd}R^{cd}-\frac{d}{4(d-1)}R^2\right)-\frac{d}{2(d-1)}RR_{ab}\right.\nonumber\\
&&\hspace{0.3cm}\left.+2R^{cd}R_{cadb}-\frac{d-2}{2(d-1)}\nabla_a \nabla_b R+\Box R_{ab}
-\frac{1}{2(d-1)}\gamma_{ab}\Box R\right\}{\cal F}(d-5).\nonumber\\
\eear

To understand the role of the BEM tensor,
let us introduce a UV cut-off for the boundary theory with $z=\mu \ne 0$.
The ultraviolet effect,
on the other hand, corresponds to the
infrared effect in AdS space
due to the so-called `UV/IR connection' \cite{Susskind:1998dq}:
$r(\mu)$ plays the role of IR regulator
in the dual bulk theory.
Thus, in light of AdS/CFT correspondence,
the renormalized AdS action
will lead to the vacuum expectation value (VEV) of the BEM.
If one rescales the BEM tensor $t_{ab}$ as
\bear\label{btt}
t_{ab}\equiv\frac{2}{\sqrt{-g}}\frac{\delta(A_{\partial B}+A_{ct})}{\delta g^{ab}}
=(\frac{l}{z})^{d-2}T_{ab}\,,
\eear
the vacuum expectation value at the boundary ($z=0$)
is given as
\cite{Myers:1999psa,de Haro:2000xn}.
\bear\label{bvev}
t_{ab}=\left\{
\begin{array}{ll}
\frac{dl^{d-1}}{16\pi G_{d+1}}g_{ab}^{(d)},&~~{\rm when}~d={\rm odd}, \\
\frac{dl^{d-1}}{16\pi G_{d+1}}
\left(g_{ab}^{(d)}+X_{ab}[g^{(n)}]\right), &~~{\rm when}~d={\rm even}\,. \\
\end{array}
\right.
\eear
It is noted that at $z=0$, $\sum_{a,b}t_{ab}g^{(0)ab}=0$.
Hence, the odd dimensional boundary is conformal.
In even dimensional boundary, $X_{ab}[g^{(n)}]$ is a function of $g^{(n)}$
with $n<d$ and reflects the conformal anomalies at the boundary.

In the presence of regulator $z=\mu \ne 0$,
the BEM tensor may hint at the effective theory at given scale
$\mu$, which may originate from high-energy effect.
In this paper, we are going to investigate the counter-term action for 4-dimensional
AdS, S-AdS and Taub-NUT-(A)dS
explicitly and find BEM tensor at 3-dimensional regulated boundary.
To do this, we employ the FG coordinates system
and the so-called `counter-term subtraction method'.
Even though, 3-dimensional trace anomaly vanishes at $z =0$,
the quantity does not vanish at $z \ne 0$ and
is expressed in terms of geometric quantity,
the boundary volume and the Ricci scalar.
Explicitly, the trace can be written as
\bear\label{Bulktr2}
\sum t_{ab}g^{ab}=\sum_{i=0}\left(\alpha_{i}\,R^{i-1}
\,\sqrt \frac{g_0}{\gamma}
+\beta_{i}\,R^{2i+1}\right),
\eear
where $g_0$ is the determinant of the metric at $z=0$,
and $\alpha_{i}$ and $\beta_{i}$ are certain constants
depending on the manifold.
The S-AdS$_4$ on the Poincare patch is the exception,
whose the trace anomaly is of the form
\bear\label{Bulktr1}
\sum t_{ab}g^{ab}=\frac{\alpha_0}{\sqrt{-\gamma}}.
\eear
It seems that this special case is due to the flat
boundary metric at any $z$.

This paper is organized as follows:
S-AdS$_4$ is considered in section \ref{sec1} and
Taub-NUT-(A)dS is considered in section \ref{sec2}.
BEM tensor is explicitly presented in FG coordinates
and its trace anomaly is shown to reduce to the form
(\ref{Bulktr2}) and (\ref{Bulktr1}).
The BEM tensor is presented in a closed-form
for the S-AdS$_4$ on a Poincare patch and for Taub-NUT-dS$_4$
when cosmological constant satisfies so-called `massless-NUT condition'
(see section \ref{sub33} for the definition).
For other types of manifolds
the BEM tensor is presented in a asymptotic series expansion of $z$.
Section \ref{sec4} is the conclusion.

\setcounter{equation}{0}
\section{S-AdS$_4$}\label{sec1}
In this section, we construct the closed form of
the transformation law from $r$ to $z$ in FG coordinate using
S-AdS$_4$ on the Poincare patch
to find the closed-form of the BEM tensor.
The trace anomaly is given  exactly as (\ref{Bulktr1}).
S-AdS$_{4}$ in the Global coordinate
are also considered. Even though, closed form of the BEM is not presented,
one can confirm that the trace anomaly satisfies (\ref{Bulktr2}).

\subsection{S-AdS$_4$ on the Poincare patch}\label{PAds4}
The metric of S-AdS$_{4}$ in the Poincare coordinate is given as
\bear\label{Pbh4}
ds^2=-f_{\rm P}(r) dt^2
+\frac{dr^2}{f_{\rm P}(r)}+
\frac{r^2}{l^2}
\left( d {x_1}^2+ d{x_2}^2\right),
\eear
where $f_{\rm P}(r)$ is given as
\bear\label{Pm}
f_{\rm P}(r)=\frac{r^2}{l^2}-\frac{m}{r},
\eear
and $m$ is a geometric mass.
The asymptotic form of BEM tensor in (\ref{fullemt}) has the form
as $r\rightarrow \infty$,
\bear\label{Pemt4series}
8\pi G_4 T_{tt}&=&\frac{m}{l r}+\frac{m l}{2r^3}
+{\cal O}(\frac{1}{r^4}),\nonumber\\
8\pi G_4 T_{ii}&=&\frac{m}{2 l r}-\frac{3 m l^3}{4 r^3}
+{\cal O}(\frac{1}{r^4}).
\eear
A conserved mass is given by the integration
over a $d-1$-surface $\Sigma$ \cite{Brown:1992br}:
\bear\label{conservedmass}
M(r)=\int_\Sigma d^{d-1}x\sqrt{\gamma}\;T_{ab}u^{a}\xi^{b}N_{lapse},
\eear
where the timelike unit normal vector $u^a$, the timelike Killing vector $\xi^{a}$, and
the square root of the lapse function $N_{lapse}$ are respectively
\bear\label{etc1}
u^a=(\frac{1}{\sqrt{f(r)}},0,\cdots),\hspace{1cm}\xi^a=(1,0,\cdots),\hspace{1cm}
N_{lapse}=\frac{1}{\sqrt{f(r)}}.
\eear
where $f(r)$ denotes the metric function in black hole solutions in (\ref{adsact}).
At $r=\infty$, the finite piece of~(\ref{conservedmass})
becomes the ADM mass $M$
\cite{Balasubramanian:1999re}
\bear\label{Pads4denergy}
M=\frac{m V_2}{8\pi G_4 l^2}\,.
\eear

One may put the system in FG coordinate and investigate
the asymptotic behavior systematically.
The transformation is given in terms of
the black hole line element
\bear\label{rint}
\int \frac{dr}{\sqrt{f_{\rm P}(r)}}=- l \int \frac{dz}{z},
\eear
whose solution is given as
\bear\label{Pads4raius}
r(z)=\frac{l(1+\bar m c_4^3 z^3)^{\frac{2}{3}}}{2^{\frac{2}{3}}c_4 z},
\eear
where $\bar{m}=m/l$.
$c_4$ is a dimensionless constant
due to the scaling degree of freedom in FG coordinate
and may be fixed $c_4=1$.
Thus, the metric in FG coordinates is written as
\bear\label{PFGbh}
ds^2&=&-A_{\rm P}(z)dt^2+B_{\rm P}(z)\left( d{x_1}^2+ d{x_2}^2\right)
+\frac{l^2}{z^2}dz^2,
\\
A_{\rm P}(z)&=&\frac{(-1+\bar{m} z^3)^2}{2^{\frac{4}{3}}
z^2(1+ \bar{m} z^3)^{\frac{2}{3}}},\qquad
B_{\rm P}(z)=\frac{(1+ \bar m z^3)^{\frac{4}{3}}}
{2^{\frac{4}{3}} z^2}.
\eear
From this, one finds the BEM tensor using the relation (\ref{btt}),
\bear\label{PAdS4zenergymoment}
8\pi G_4 t_{tt}&=&\frac{2^{\frac{2}{3}}\bar{m}
(-1+\bar{m} z^3)^2}{(1+\bar{m} z^3)^{\frac{5}{3}}},\\
8\pi G_4 t_{ii}&=&-\frac{\bar{m}(1+\bar{m} z^3)^{\frac{1}{3}}(1+2\bar{m}z^3)}
{\sqrt[3]{2} \;(-1+\bar{m} z^3)}.
\eear
As a check, one may integrate  (\ref{conservedmass})
using the explicit form of (\ref{PAdS4zenergymoment}) in FG coordinate
to find the ADM mass in (\ref{Pads4denergy}).

The BEM tensor in (\ref{PAdS4zenergymoment})
results in the trace
\bear\label{bsp4tr}
\sum t_{ab} g^{ab}=\frac{12\bar{m}^2l^2z^3}{8\pi G_4(1-\bar{m}^2z^6)}
=\frac{l^2}{\sqrt{A_{\rm P} {B}_{\rm P}^2}}
\left(\frac{\bar{m}^2}{8\pi{{G_4}}}\right) \,.
\eear
This shows that the trace anomaly is inversely proportional
to the boundary volume
as claimed in (\ref{Bulktr1})
where $\alpha_0= {\bar{m}^2}/{8\pi{{G_4}}}$
and other coefficients vanishes.
The expression (\ref{bsp4tr})
is also checked using the relation
with the extrinsic curvature (\ref{fullemt}, \ref{emts})
\bear
\sum t_{ab} g^{ab}=\frac{l^3}{8\pi G_4z^3}\left({-2\Theta}-\frac{6}{l}\right),
\eear
and the explicit expression of $\Theta$ in FG coordinates.

\subsection{AdS$_4$ on the Global patch}
Since AdS$_4$ in the Poincare coordinate is trivial,
we consider only AdS$_4$ on the Global patch
\bear\label{PureGbh4}
ds^2=-\left(1+\frac{r^2}{l^2}\right)dt^2
+\frac{dr^2}{1+\frac{r^2}{l^2}}+r^2 d\Omega_2,
\eear
where $d\Omega_2$ denotes the 2-sphere
metric
$d\Omega_2=d\theta^2 + \sin^2\theta d\phi^2$.
The asymptotic form of BEM tensor is
\bear\label{PureGAdS4renergymoment}
8\pi G_4T_{tt}&=&\frac{l}{4r^2}+{\cal O}(\frac{1}{r^4}),\nonumber\\
8\pi G_4T_{\theta\theta}&=&\frac{l^3}{4r^2}+{\cal O}(\frac{1}{r^4})
\,,\qquad
T_{\phi\phi} =T_{\theta\theta} \times\sin^2\theta,
\eear
and the ADM mass vanishes, $M=0$.
Using the coordinate transformation,
$ r(z)=-\frac{l(z^2-1)}{2z}$
one has the FG metric
\bear\label{PureGFGbh}
ds^2&=&-{\cal A}_{\rm G}(z)dt^2+{\cal B}_{\rm G}(z)d\Omega_2^2
+\frac{l^2}{z^2}dz^2,
\\
\label{PureGobal4z}
{\cal A}_{\rm G}(z)&=&\frac{(2^{2/3}z^2+1)^2}{4(2^{2/3})z^2}\,,\qquad
{\cal B}_{\rm G}(z)=\frac{l^2(2^{1/3}z-1)^2(2^{1/3}z+1)^2}{4(2^{2/3})z^2},
\eear
and the BEM tensor
\bear\label{PureGAdS4zenergymoment}
8\pi G_4t_{tt}=\frac{2^{1/3}z(1+2^{2/3}z^2)^2}{(-1+2^{2/3}z^2)^2}
\,,
8\pi G_4t_{\theta\theta}=\frac{2^{1/3}zl^2(1-2^{2/3}z^2)}{1+2^{2/3}z^2}
\,,
t_{\phi\phi}= t_{\theta\theta}\sin^2\theta \,.
\eear
Note that at $z=0$, $t_{tt}=0$ leads to $M=0$,
consistent with (\ref{Pads4denergy}),
S-AdS$_4$ result with  $m=0$.

The close form of  the trace of BEM  is given as
\bear
\sum t_{ab}g^{ab}=\frac{l^3}{8\pi G_4 z^3}
\left({-2\Theta}-\frac{6}{l}-\frac{l}{2}R\right)
=\frac{2^{1/3}z (1-3 (2^{2/3})z^2)l^2}{8\pi G_4(1+2^{2/3}z^2)(-1+2^{2/3}z^2)^2}\,.
\eear
This reduces to the form suggested in (\ref{Bulktr2})
\bear\label{pureads4tr}
\sum t_{ab}g^{ab}
=\frac{\sqrt{g_0}}{\sqrt{{\cal A}_{\rm G}{\cal B}_{\rm G}^2}}
\left(\frac{l^2}{\pi G_4R}
-\frac{l^4}{4\pi G_4}\right) \,,
\eear
where $\alpha_0= l^2/\pi G_4$
and $\alpha_1= -l^4/4\pi G_4$
and other coefficients vanish.

\subsection{S-AdS$_4$ on the Global patch}\label{S-AdS4}
The metric of S-AdS$_4$ on the Global patch is written as
\bear\label{Gbh4}
ds^2&=&-f_{\rm G}(r)dt^2
+\frac{dr^2}{f_{\rm G}(r)}+r^2 d\Omega_2,
\\
\label{omega2}
f_{\rm G}(r)&=& \left(1+\frac{r^2}{l^2}-\frac{m}{r}\right)\,.
\eear
The BEM tensor is given asymptotically as \cite{Balasubramanian:1999re}
\bear\label{GAdS4zenergymoment}
8\pi G_4T_{tt}&=&\frac{m}{lr}+\frac{1}{4lr^2}+\frac{ml}{2r^3}+{\cal O}(\frac{1}{r^4}),\nonumber\\
8\pi G_4T_{\theta\theta}&=&\frac{ml}{2r}+\frac{l^3}{4r^2}-\frac{3ml^3}{4r^3}+{\cal O}(\frac{1}{r^4})
\,,\quad  T_{\phi\phi} = T_{\theta\theta}\times\sin^2\theta,
\eear
and the ADM mass (\ref{conservedmass}) is
\bear\label{Gads4denergy}
M=\frac{m}{2 G_4}.
\eear

On the other hand, the FG metric
\bear\label{GFGbh}
ds^2=-A_{\rm G}(z)dt^2+B_{\rm G}(z)d\Omega_2^2
+\frac{l^2}{z^2}dz^2,
\eear
is given in term of the elliptic function. Thus, we may expand
$ 1/{\sqrt{f_{\rm G}(r)}}$ in power series of $1/r$ to have
\bear\label{frexpansion}
\frac{1}{\sqrt{f_{\rm G}(r)}}= l\left(\frac{1}{r}-\frac{l}{2r^3}+\frac{m l^2}{r^4}\right)
+{\cal O}(\frac{1}{r^5}),
\eear
and the relation (\ref{rint}) shows
\bear\label{ads4radius}
r(z)=l\left[\frac{5^{1/3}}{z}-\frac{z}{2(5^{1/3})}+\frac{2\bar{m} z^2}{3(5^{2/3})}
-\frac{3 z^3}{20}+{\cal O}(z^4)\right],
\eear
and
\bear\label{Gobal4z}
A_{\rm G}(z)&=&\frac{1}{z^2}\left(\frac{5^{2/3}}{4}+\frac{z^2}{2}
-\frac{4\bar{m}z^3}{3(5^{1/3})}-\frac{z^4}{2(5^{2/3})}
-\frac{8\bar{m}z^5}{15}+{\cal O}(z^6)\right),\nonumber\\
B_{\rm G}(z)&=&\frac{l^2}{z^2}\left(\frac{5^{2/3}}{4}-\frac{z^2}{2}
+\frac{3\bar{m} z^3}{3(5^{1/3})}
-\frac{z^4}{2(5^{2/3})}-\frac{2\bar{m} z^5}{15}+{\cal O}(z^6)\right).
\eear
The BEM tensor in FG coordinate is given as
\bear\label{GAdS4energymomseries}
8\pi G_4 t_{tt}&=&
\frac{2\bar{m}}{5^{1/3}}-\frac{2z}{5^{2/3}}
+\frac{14\bar{m}z^2}{15}+{\cal O}(z^3),\nonumber\\
8\pi G_4 t_{\theta\theta}&=&l^2\left(\frac{2\bar{m}}{5^{1/3}}+\frac{4z}{5^{2/3}}
-\frac{11\bar{m}z^2}{15}+{\cal O}(z^3)\right)
\,,\quad
t_{\phi\phi}= t_{\theta\theta} \times\sin^2\theta\,.
\eear

From the BEM tensor
one can put the series expansion of the
trace into the form
\bear\label{rssads4g}
\sum t_{ab}g^{ab}
&=&\frac{l^3}{8\pi G_4 z^3}\left({-2\Theta}-\frac{6}{l}
-\frac{l}{2}R\right)\nonumber\\
&=&\frac{\sqrt{g_0}}{R\sqrt{A_{\rm G}B_{\rm G}^2}}
\left(\frac{l^2}{\pi G_4}\right)
+\left(-\frac{\bar{m}{l^4}}{20\pi G_4}\right)\,R+{\cal O}(z^3),
\eear
which agrees with (\ref{pureads4tr}) when $m=0$.
This result shows that
the leading term is inversely proportional to the boundary volume
and the next leading term is proportional to the Ricci scalar squared,
which corresponds to
$\alpha_0 =l^2/\pi G_4$
and $\beta_0=-\bar{m}{l^4}/20\pi G_4$
in  (\ref{Bulktr2}).

\setcounter{equation}{0}
\section{Taub-NUT-(A)dS}\label{sec2}
In this section, we consider the  Taub-NUT-AdS$_{4}$ on
the Poincare and on the Global patch.
The BEM tensor and its trace anomaly
is given asymptotically
in terms of FG coordinates.
It is noted that
if one considers Taub-NUT-dS metric
with NUT mass zero condition,
then BEM tensor and its trace anomaly
is given in a rational function.
We will investigate this possibility
in subsection \ref{sub33} Taub-NUT-dS
on the Global patch.

\subsection{Taub-NUT-AdS$_4$ on the Poincare patch}
\label{PTaubNUTsubsection}
The metric of Taub-NUT-AdS on the Poincar\'e patch is given as
\cite{Emparan:1999pm,Chamblin:1998pz}:
\bear\label{Ptaubads4}
&&ds^2 = -g_{\rm P}(r) \left(dt +
\frac{n}{l^2}(x_1 dx_2-x_2dx_1)\right)^2
+\frac{dr^2}{g_{\rm P}(r)}
+\frac{r^2+n^2}{l^2}
\left( dx_1^2+ dx_2^2\right),
\nn \\
&& g_{\rm P}(r)=\frac{r^4+6n^2r^2-3n^4}{l^2 (r^2+n^2)}-\frac{mr}{r^2+n^2}.
\eear
The BEM tensor is given asymptotically as
\bear\label{PTaubAdS4energymoment}
8\pi G_4 T_{tt}&=&\frac{m}{l r}+\frac{105n^4}{4l^3r^2}-\frac{5mn^2}{2lr^3}
+\cdots ,\nonumber\\
8\pi G_4 T_{tx_1}&=&-\frac{m n x_2}{l^3 r}+\frac{105 n^5 x_2}{4 l^5 r^2}+
\frac{5mn^3x_2}{2 l^3 r^3}+\cdots ,\nonumber\\
8\pi G_4 T_{tx_2}&=&\frac{m n x_1}{l^3 r}+\frac{105 n^5 x_1}{4 l^5 r^2}
-\frac{5mn^3x_1}{2 l^3 r^3}+\cdots ,\nonumber\\
8\pi G_4 T_{x_1x_1}&=&\frac{m(l^4 + 2n^2 x_2^2)}{2 l^5 r}
+\frac{n^4(63l^4+105n^2 x_2^2)}{4l^7 r^2}-\frac{mn^2(21l^4+10n^2x_2^2)}{4 l^5 r^3}
+\cdots ,\nonumber\\
8\pi G_4 T_{x_2x_2}&=&\frac{m(l^4 + 2n^2 x_1^2)}{2 l^5 r}
+\frac{n^4(63l^4+105n^2 x_1^2)}{4l^7 r^2}-\frac{mn^2(21l^4+10n^2x_1^2)}{4 l^5 r^3}
+\cdots ,\nonumber\\
8\pi G_4 T_{x_1x_2}&=&\frac{m n^2 x_1 x_2}{l^5 r}-\frac{105 n^6 x_1 x_2}{4 l^7 r^2}+
\frac{5mn^4 x_1 x_2}{2 l^5 r^3}+\cdots ,
\eear
where $\cdots$ denotes the terms of ${\cal O}({1}/{r^4})$.
The ADM mass is given as
\bear\label{PTaubAds4denergy}
M=\frac{m V_2}{8\pi G_4 l^2},
\eear
where $V_2$ refers to the 2-dimensional volume.
From (\ref{rint}) $r(z)$ is written in powers of $z$
\bear\label{PTaubAdS4radisus}
r(z)=l\left(\frac{1}{z}-\frac{5\bar{n}^2 z}{4}+\frac{\bar{m} z^2}{6}
-\frac{75 \bar{n}^4 z^3}{32}+ {\cal O}(z^4)\right),
\eear
where $\bar{n}=n/l$, and the FG metric is written as
\bear\label{PTaubAdS4FG}
ds^2&=&-E_{\rm P}(z)\left(dt + \frac{n}{l^2}(x_1dx_2-x_2dx_1)\right)^2
+F_{\rm P}
\left( d{x_1}^2+ d{x_2}^2\right)
+\frac{l^2}{z^2}dz^2,
\\
\label{PTaubAdS4CD}
E_{\rm P}(z)&=&\frac{1}{z^2}\left(1+\frac{5\bar{n}^2z^2}{2}-\frac{2\bar{m}z^3}{3}
-\frac{89\bar{n}^4z^4}{8}
-\frac{2\bar{m}\bar{n}^2z^5}{3}+{\cal O}(z^6)\right), \nonumber\\
F_{\rm P}(z)&=&\frac{l^2}{z^2}\left(1-\frac{3 \bar{n}^2z^2}{2}+\frac{\bar{m} z^3}{3}
-\frac{25 \bar{n}^4 z^4}{8}-\frac{5 \bar{m} \bar{n}^2 z^5}{12}+{\cal O}(z^6)\right).
\eear
The BEM tensor is
\bear
8\pi G_4 t_{tt}&=&
{\bar{m}}-\frac{\bar{n}^4 z}{2}-\frac{13\bar{m}  \bar{n}^2 z^2}{12}
+{\cal O}(z^3),\nonumber\\
8\pi G_4 t_{tx_1}&=&\
-{\bar{m} \bar{n} \bar{x}_2}+\frac{\bar{n}^5 \bar{x}_2 z}{2}+
\frac{13\bar{m}\bar{n}^3\bar{x}_2z^2}{12}+{\cal O}(z^3),\nonumber\\
8\pi G_4 t_{tx_2}&=&
{\bar{m}  \bar{n} \bar{x}_1}-\frac{\bar{n}^5 \bar{x}_1 z}{2}-
\frac{13\bar{m}\bar{n}^3\bar{x}_1z^2}{12}+{\cal O}(z^3),\nonumber\\
8\pi G_4 t_{{x}_1 {x}_1}&=&
\frac{\bar{m} (1+2\bar{n}^2\bar{x}_2^2)}{2}
+\frac{\bar{n}^4(85-\bar{n}^2 \bar{x}_2^2)z}{2}
-\frac{\bar{m}\bar{n}^2(115+26\bar{n}^2\bar{x}_2^2)z^2}{24}
+{\cal O}(z^3),\nonumber\\
8\pi G_4 t_{x_2x_2}&=&
\frac{\bar{m} (1+2\bar{n}^2\bar{x}_1^2)}{2}
+\frac{\bar{n}^4(85-\bar{n}^2 
(115+26\bar{n}^2\bar{x}_1^2)z^2}{24}
+{\cal O}(z^3),\nonumber\\
8\pi G_4 t_{x_1\,x_2 }&=&
-{\bar{m} \bar{n}^2 \bar{x}_1 \bar{x}_2}
+\frac{\bar{n}^6 \bar{x}_1 \bar{x}_2\, z}{2}
+\frac{13\bar{m}\bar{n}^4\bar{x}_1 \bar{x}_2\,z^2}{12}+{\cal O}(z^3)\,,
\eear
where $\bar{x}_1=x_1/l$ and $\bar{x}_2=x_2/l$.
$t_{tt}=\frac{\bar{m}}{8\pi G_4}$
gives the VEV in boundary CFT and
provides the mass in (\ref{PTaubAds4denergy}).

The trace anomaly has the form
\bear
\sum t_{ab}g^{ab}&=&\frac{1}{R\sqrt{E_{\rm P}F_{\rm P}^2}}
\left(\frac{171 \bar{n}^6 l^2}{8 \pi G_4 }\right)
+\left(-\frac{9\bar{m} l^4}{32 \pi G_4}\right)\,R
+{\cal O}(z^3),
\eear
which reproduces (\ref{Bulktr2}),
where
$\alpha_0={171 \bar{n}^6}l^2/{8 \pi G_4 }$
and $\beta_0=-9\bar{m} l^4/{32 \pi G_4}$.

\subsection{Taub-NUT-AdS$_4$ on the global patch}\label{sub32}
The metric of Taub-NUT-AdS$_4$ on the global patch
is given by \cite{Awad:2000gg,Clarkson:2005aa}
\bear\label{Gtaubbh4}
ds^2&=&-g_{\rm G}(r)(dt+2n \cos\theta d\phi)^2+\frac{dr^2}{g_{\rm G}(r)}
+(r^2+n^2)d\Omega_2,
\\
\label{g4}
g_{\rm G}(r)&=&
\frac{l^2 r^2-n^2 l^2+r^4+6n^2r^2-3n^4}{l^2 (r^2+n^2)}
-\frac{m_{\rm G}r}{r^2+n^2}.
\eear
The EM tensor is
\bear
8\pi G_4 T_{tt}&=&\frac{m}{l r}+\frac{l^4+30 n^2 l^2+105n^4}{4l^3r^2}+\frac{m(l^2-5n^2)}{2lr^3}
+{\cal O}(\frac{1}{r^4}),\nonumber\\
8\pi G_4 T_{t\phi}&=&
\cos\theta \left(
\frac{m n (6 l-4) }{l^2 r}
+\frac{n(l^4+30 n^2 l^2+105 n^4)}{2 l^3 r^2}
+\frac{m n (l^2-5n^2)}{l r^3}+{\cal O}(\frac{1}{r^4})
\right),
\nonumber\\
8\pi G_4 T_{\phi\phi}&=&\frac{m (8n^2+(l^2-8 n^2)\sin^2\theta)}{2l r}\nonumber\\
&&+\frac{4n^2(l^4+30n^2l^2+105n^4)+(l^2+4n^2)(l^4+12 n^2l^2-105n^4)\sin^2\theta}{4 l^3 r^2}\nonumber\\
&&-\frac{m(-8n^2(l^2-5n^2)+(3l^4+29n^2l^2-40n^4)\sin^2\theta)}{4 l r^3}
+{\cal O}(\frac{1}{r^4}),\nonumber\\
8\pi G_4 T_{\theta\theta}
&=& \frac{m l}r + \frac{l^4 + 20n^2 l^2}{4 l r^2}
+ \frac{ml(3 l^2 + 21 n^2) }{2 r^3}
+{\cal O}(\frac{1}{r^4})\,,
\eear
and the ADM  mass (\ref{conservedmass}) is
\bear\label{taub4denergy}
M=\frac{m}{2G_4}.
\eear

Using the asymptotic form of $r(z)$
\bear\label{Gtaub4radius}
r(z)=l\left(1-\frac{(1+5\bar{n}^2)z}{4}
+\frac{\bar{m}z^2}{6}
-\frac{(3+30\bar{n}^2+75\bar{n}^4)z^3}{32}
+{\cal O}(z^4)\right),
\eear
one has the metric in the FG
\bear\label{GTaubAdS4}
ds^2&=&-{\cal E}_{\rm G}(z)(dt+2n \cos\theta d\phi)^2
+{\cal F}_{\rm G}(z)d\Omega_2 +\frac{l^2}{z^2}dz^2,
\\
{\cal E}_{\rm G}(z)&=&\frac{1}{z^2}\left(1
+\frac{(1+5\bar{n}^2)z^2}{2}-\frac{2\bar{m}z^3}{3}
+\frac{(1+26\bar{n}^2+89\bar{n}^4)z^4}{8}
+ \frac{\bar{m}(1+2\bar{n}^2)z^5}{3}+{\cal O}(z^6)\right),
\nonumber\\
{\cal F}_{\rm G}(z)&=&\frac{l^2}{z^2}\left(1-\frac{(1+3\bar{n}^2)z^2}{2}
+\frac{2\bar{m} z^3}{3}
-\frac{(1+5\bar{n}^2)^2z^4}{8}-\frac{\bar{m} (1+5\bar{n}^2)z^5}{6}
+{\cal O}(z^6)\right)\,.
\nonumber
\eear
The BEM tensor in FG coordinates is asymptotically given as
\bear\label{ttcom44d}
8\pi G_4 t_{tt}&=&2\bar{m}-\frac{(4\bar{n}^2+\bar{n}^4+1)}{2}z+
\frac{\bar{m}(-13\bar{n}^2+7)}{6}z^2 +{\cal O}(z^3),\nonumber\\
8\pi G_4 t_{t\phi}&=&l\left({2\bar{m}\bar{n}\cos\theta}
-{\bar{n}(1+4 \bar{n}^2 + \bar{n}^4)z}
+\frac{\bar{m}\bar{n}(7-13\bar{n}^2)}{6}z^2+{\cal O}(z^3)\right),\nonumber\\
8\pi G_4 t_{\phi\phi}&=& l^2\left(\frac{\bar{m}(8\bar{n}^2
+(1-8\bar{n}^2)\sin^2\theta)}{2}\right.
\nonumber\\
&&+\frac{[-4\bar{n}^2(1+4\bar{n}^2+\bar{n}^4)+(1+4\bar{n}^2)
(2+25\bar{n}^2+\bar{n}^4)\sin^2\theta]z}
{4}\nonumber\\
&&\left.-\frac{\bar{m}[-56\bar{n}^2+104\bar{n}^4
+(11+171\bar{n}^2-104\bar{n}^4)\sin^2\theta]z^2}{24}+{\cal O}(z^3)\right),\nonumber\\
8\pi G_4 t_{\theta\theta}&=&l^2\left(\frac{\bar{m}}{2}
+\frac{(2+29\bar{n}^2+85\bar{n}^4)z}{2}
-\frac{\bar{m}(11+115\bar{n}^2)z^2}{24}+{\cal O}(z^3)\right).
\eear
This BEM tensor leads to  the VEV in boundary CFT
$ t_{tt}={\bar{m}}/{(8\pi G_4)}$,
which reproduces the mass (\ref{taub4denergy}).

This shows that the trace of BEM  has the form
\bear
\sum t_{ab}g^{ab}&=&
\frac{\sqrt{g_0}}{R\sqrt{{\cal E}_{\rm G}{\cal F}_{\rm G}^2}}
\left(\frac{(1+\bar{n}^2)l^2(5+62\bar{n}^2+171\bar{n}^4)}
{8\pi G_4}\right)\nonumber\\
&&+\left(-\frac{\bar{m} l^4 (1+9\bar{n}^2)}
{16\pi G_4 (1+\bar{n}^2)}\right)\,R
+{\cal O}(z^3),
\eear
which results in
$\alpha_0= (1+\bar{n}^2)l^2(5+62\bar{n}^2+171\bar{n}^4)
/8\pi G_4$
and
$\beta_0=-\bar{m} l^4 (1+9\bar{n}^2)/16\pi G_4 (1+\bar{n}^2)$
in (\ref{Bulktr2}).

\subsection{Massless Taub-NUT-dS}\label{sub33}

Taub-NUT-dS metric is obtained
from the Taub-NUT-AdS metric
by replacing $l^2 \to -l^2$.
On the global patch one has
\cite{Clarkson:2003wa,Clarkson:2003kt,Mann:2004mi,Clarkson:2005aa}
\bear\label{tnds4m}
&&ds^2
= g_{\rm G, dS}(r)(dt+2n \cos\theta d\phi)^2
-\frac{dr^2}{g_{\rm G, dS}(r)}
+(r^2+n^2)d\Omega_2,
\\
&& g_{\rm G, dS}(r)
=\frac{-l^2 r^2+n^2 l^2+r^4+6n^2r^2-3n^4}{l^2 (r^2+n^2)}
+\frac{m_{\rm G, dS}\, r}{r^2+n^2}
\nn.
\eear
Due to the signature change of the metric
compared with that of AdS, one
may view this dS metric
as the one describing
the region outside of the cosmological event horizon.
The future infinity ($r \to \infty$)
corresponds to the asymptotic boundary
we are paying attention to.

Let us consider the metric with
$n^2 = l^2/4$ and $m_{\rm G, dS} =0$.
In this case, the metric is simplified
\bear
g_{\rm G,dS} (r) &=&
\frac{r^2 + l^2/4}{l^2}\,.
\eear
Using
$ r(z)=l\left(\frac{1-l^2 z^2/4}{2z}\right)$
one finds the metric in FG coordinates
\bear\label{GTaubdS4}
&&ds^2={\cal E}_{\rm G, dS}^{(1)}(z)(dt+2n \cos\theta d\phi)^2
+{\cal F}_{\rm G, dS}^{(1)}(z)d\Omega_2-\frac{l^2}{z^2}dz^2,
\\
&&{\cal E}_{\rm G, dS}^{(1)}(z)=
\left(\frac{4+z^2}{8z}\right)^2\,,\qquad
{\cal F}_{\rm G, dS}^{(1)}(z)=
l^2\left(\frac{4+z^2}{8z}\right)^2.
\nn
\eear

It is noted that the conditions,
$n^2 = l^2/4$ and $m_{\rm G, dS} =0$
are not independent in the
Wick-rotated Euclidean section.
To see this, one Wick-rotates
the time ($(t\rightarrow i t_E $)
and NUT charge ($ n\rightarrow i N)$ simultaneously
from (\ref{tnds4m})
and obtains the metric
\bear
&&
-g_{\rm G, dS}^{E}(r)(dT+2N \cos\theta d\phi)^2-\frac{dr^2}{g_{\rm G, dS}^{E}(r)}
+(r^2-N^2)d\Omega_2,
\\
\nn
&& g_{\rm G, dS}^{E}(r)=\frac{-l^2 r^2-N^2 l^2+r^4-6N^2r^2-3N^4}{l^2 (r^2-N^2)}
+\frac{m_{\rm G, dS}\,r}{r^2-N^2}.
\eear
Requiring  $g_{\rm G, dS}(r)|_{r=N}=0$
one has the relation
\bear
m_{\rm G, dS}=\frac{2N(l^2+4N^2)}{l^2}\,.
\eear
Thus, if one puts $n^2 = -N^2 =l^2/4$,
then $m_{\rm G, dS}=0$.
Because of this relation, we will call
$n^2 = l^2/4$ the massless-NUT condition.
In addition,
if one requires the regularity of the metric
at the cosmological event horizon ($r=N$),
one finds~\cite{Astefanesei:2004ji}
\bear
\frac{4\pi}{ \left|
{\partial g_{\rm G, dS}^{E}(r)}/{\partial r}
\right|} (r=N)
=8\pi  N,
\eear
which is interpreted as the inverse temperature
at the cosmological event horizon.

The BEM tensor is given from the minimal
counter-term subtraction (\ref{fullemt})
\cite{Clarkson:2003wa,Clarkson:2003kt,Mann:2004mi}
\bear\label{chPemt4}
T_{ab}=\frac{1}{8\pi G_{4}}\left[\Theta_{ab}-\Theta\gamma_{ab}
-\frac{2}{l}\gamma_{ab}-l(R_{ab}-\frac{1}{2}\gamma_{ab}R)\right],
\eear
where the last term
switches the sign to negative
from the one in AdS case.
In $r$-coordinate system, the BEM tensor
is given asymptotically as
\bear\label{GTaubdS4energymoment}
8\pi G_4 T_{tt}&=&
-\frac{n}{32r^2}+\frac{n^3}{64r^4}+{\cal O}(\frac{1}{r^6}),\nonumber\\
8\pi G_4 T_{t\phi}&=&-\frac{n^2\cos\theta}{r^2}+\frac{n^4\cos\theta}{32r^4}
+{\cal O}(\frac{1}{r^6}),\nonumber\\
8\pi G_4 T_{\phi\phi}&=&
8\pi G_4 T_{\theta\theta} = -\frac{n^3}{r^2}+\frac{n^5}{16r^4}
+{\cal O}(\frac{1}{r^6}),
\eear
and the ADM mass vanishes $M=0$.
In $z$-coordinates,  the BEM is given
in a closed form
\bear\label{GTaubdS4energymomentz}
8\pi G_4 t_{tt}&=&-\frac{\bar{n} z}{8},\nonumber\\
8\pi G_4 t_{t\phi}&=&-\frac{l\bar{n}^2z\cos\theta}{4},\nonumber\\
8\pi G_4 t_{\phi\phi}&=&
8\pi G_4 t_{\theta\theta}
=-\frac{l \bar{n}^3 z}{2}\,.
\eear
The trace of BEM tensor is obtained as
\bear
\sum t_{ab}g^{ab}&=&\frac{6\bar{n}^6 l^2 z}
{\pi G_4(1+\bar{n}^2 z^2)^2}\nonumber\\
&=&\frac{\sqrt{g_0}}{R\sqrt{{\cal E}_{\rm G, dS}^{(1)}({\cal F}_{\rm G, dS}^{(1)})^2}}
\left(\frac{9\bar{n}^5l^2}{4\pi G_4}
-\frac{3\bar{n}^7l^2}{4\pi G_4}\right).
\eear
This shows that
$\alpha_0=9\bar{n}^5l^2/4\pi G_4$,
$\alpha_1= -3\bar{n}^7l^2/4\pi G_4$
and others vanish
in (\ref{Bulktr2}).

It is remarked that
the massless-NUT condition can also be
imposed on the Taub-NUT-dS metric
in arbitrary even dimensions.
The Taub-NUT-dS
with $k \,S_2 $ spheres is given
in Euclidian section
as \cite{Clarkson:2005aa}
\bear\label{gendsmet}
&&ds^2=-g_{\rm G, dS}^{E,(k)}(r)\left[dT+2N \sum_{i=1}^{k}\cos(\theta_i) d\phi_i\right]^2
-\frac{dr^2}{g_{\rm G, dS}^{E,(k)}(r)}
+(r^2-N^2)\sum_{i=1}^{k}d \Omega_2 ^{(i)},
\\
&&g_{\rm G, dS}^{E,(k)}(r)=\frac{r}{(r^2-N^2)^{k}}\int^r
\left[\frac{(2k+1)(s^2-N^2)^{k+1}}{l^2 s^2}-\frac{(s^2-N^2)^k}{s^2}\right]ds
+\frac{m_{\rm G}^{(k)} r}{(r^2-N^2)^k},
\nn
\eear
where $d\Omega_2 ^{(i)} $ refers to the $i$-th unit sphere measure.
In this manifold, The NUT mass is given as
\bear
m_{\rm G}^{(k)}=\Gamma\left(\frac{3-2k}{2}\right)
\Gamma(k+1)\frac{N^{2k-1}[l^2+(2k+2)N^2]}{2\sqrt{\pi}l^2(2k-1)},
\eear
where $\Gamma(k)$ is the gamma function.
Massless-NUT condition is
\bear\label{gerNUTmasszeroc}
l^2=-(2k+2)N^2=(2k+2)n^2,
\eear
and the metric in Minowski section
with this massless-NUT condition
is written as
\bear
&&ds^2=g_{\rm G, dS}^{(k)}(r)\left[dt+2n \sum_{i=1}^{k}\cos(\theta_i) d\phi_i\right]^2
-\frac{dr^2}{g_{\rm G, dS}^{(k)}(r)}
+(r^2+n^2)\sum_{i=1}^{k}d\Omega_2^{(i)},\\
\nn
&&g_{\rm G, dS}^{(k)}(r)=\frac{r^2+n^2}{(2k+2)n^2}\,.
\eear
The metric is compactly written in FG coordinates
\bear\label{genGTaubdS2k}
&&ds^2={\cal E}_{\rm G, dS}^{(k)}(z)\left[dt+2n \sum_{i=1}^{k}\cos(\theta_i) d\phi_i\right]^2
+{\cal F}_{\rm G, dS}^{(k)}(z)\sum_{i=1}^{k}d\Omega_2^{(i)}
-\frac{l^2}{z^2}dz^2,
\\ \nn
&&{\cal E}_{\rm G, dS}^{(k)}(z)=
\left(\frac{2k+2+z^2}{4(k+1)z}\right)^2\,,\qquad
{\cal F}_{\rm G, dS}^{(k)}(z)
=l^2\left(\frac{2k+2+z^2}{4(k+1)z}\right)^2,
\eear
where $
r(z)=l\left(\frac{1-n^2z^2}{2z}\right)$
is used.

\section{Conclusion and discussion}\label{sec4}

Employing the counter-term subtraction method,
we have investigated the boundary energy momentum
tensor and its trace anomaly
for the various cases of metric, including
AdS$_4$, S-AdS$_4$ and Taub-NUT-(A)dS$_4$.
Using the asymptotic expansion in FG coordinates,
we find that the trace of the BEM tensor
has a special form (\ref{Bulktr2})
in general  (and (\ref{Bulktr1}) for
S-AdS$_4$ on the Poincar\'e metric only).

The 3-dimensional holographic trace anomaly
vanishes at $r\to \infty$ (or
$z \to 0$) boundary. However,
the trace is non-vanishing
when $r$ is finite,
which corresponds to the UV cut-off
at the boundary field theory point of view.
Thus, if the AdS/CFT duality holds
in this case,
these anomalies can be interpreted
as the breaking of conformal symmetry
of the 3-dimensional field theory.
Thus, it will be interesting to find
the effective field theory
whose trace anomaly
matches with the power correction
in $z$ as the high-energy effect.

Finally, it is noted that a different regularized action
in AdS$_4$ has been
obtained in~\cite{Aros:1999id,Olea:2005gb}
where Gauss-Bonnet action with topological term
is chosen instead
of the boundary action (\ref{totadsact}),
assuming that the Riemann
curvature tensor is given as a constant negative curvature at the
boundary.
This action allows anomaly free conserved quantity at the
boundary.
Thus, one may
naively expect that (\ref{Bulktr2}) and (\ref{Bulktr1})
originates from the topological term in~\cite{Olea:2005gb}.
However, it turns out that
the 3-dimensional trace anomaly does not lead to the
boundary action.
Thus, the difference seems to arise due to
the boundary condition
imposed in \cite{Olea:2005gb}
and the Dirichlet boundary condition
employed in this work.
The two different results allude that the AdS/CFT
correspondence is sensitive to the boundary condition
imposed on the CFT boundary.
\section*{Acknowledgements}
We are grateful to Siyoung Nam, Chanyong Park, Mu-In Park, and Ho-Ung Yee
for informative discussion.
This work was supported by the BK 21 project of the Ministry of Education and Human Resources Development in Korea
and through the Center for Quantum Spacetime (CQUeST)
of Sogang University with grant (R11-2005-021).

\end{document}